# The Prevention of Singularities Inside Black Holes

Julian B. Wilson
Department of Physics
University of California, Riverside
Riverside, California 92521

**Abstract**
In this paper the currently held view that the endpoint of gravitational collapse is a singularity is refuted.  A quantum mechanical calculation is done showing that spin 0 and spin ½ particles inside a black hole's schwarzschild radius aren't confined to an infinitesimal point but form bound state orbits.  As with the case of electric collapse, if an electron cannot spiral into a nucleus, then neither does this happen in plasma consisting of many electrons and nuclei.  Showing that after undergoing gravitational collapse, like plasma, matter will not contract to an infinitesimal point.  That quantum effects prevents singularity formation.

## Introduction

It is currently believed that gravitational collapse ends with matter being confined to an infinitesimal point of infinite mass-energy density and space-time curvature (a singularity).  This is predicted when the problem is analyzed classically.  Many found this solution to be unacceptable since this would violate the uncertainty principle, and felt that quantum effects must come into play to prevent the singularity from forming [1].  These feelings were buoyed by the fact that an identical problem existed in Electricity and Magnetism, the paradox of electric collapse that existed early in the previous century.

Analyzed classically an electron orbiting a nucleus would radiate, losing its energy, and spiral into the nucleus in ~$10^{-17}$s.  This predicted atoms to be unstable, contrary to observation.  It took a quantum mechanical analysis to show that electrons under the influence of a nucleus' electric field formed bound states, showing that quantum effects prevent the electron from spiraling into the nucleus.  The problem of electric collapse was solved using 1$^{st}$ quantization, by quantizing particle motion in a classical field.  Although it was known that the electric field too needed to be quantized (known as 2$^{nd}$ quantization), such effects only contributed to small corrections at these energies.  The quantum effects from 1$^{st}$ quantization become apparent at energies of order ~1eV.  The effects of 2$^{nd}$ quantization of the electric field do not begin to dominate until energies are of order ~1MeV (with $e^+/e^-$ annihilation).

Those who believed quantum effects prevented singularities from forming inside black holes, however, felt that one must first quantize the gravitational field [1] ($2^{nd}$ quantization), even though these effects don't become apparent until energies are of order ~100GeV [2]. And a possible $1^{st}$ quantization resolution to singularity formation hadn't been sought.

This paper presents such a resolution. Solving the problem quantum mechanically for spin 0 and spin ½ particles inside the schwarzschild radius ($r_s$) of a central black hole. It will be shown that bound state wave functions do exist. Proving that after undergoing gravitational collapse, matter will not contract to an infinitesimal point. That quantum effects prevent singularities from forming.

## The Hydrogen Atom

Before applying quantum mechanics to the problem of gravitational collapse it would be insightful if one first reviews how quantum mechanics was used to solve the problem of electric collapse. This is the famous hydrogen atom wave function derivation. One first starts with the Schrödinger equation [3], where the Coulombic potential is used for the electromagnectic interaction between the proton and the orbiting electron. After performing separation of variables one arrives at a differential equation for the radial part of the wave function. Instead of attempting to find the solution in terms of elementary functions, a property of analytic functions is used. If the solution to a differential equation is finite at it's critical points, then it is finite between them [4]. So for the hydrogen atom, the solution to the radial part of the wave function is found by first looking at differential equation in the limit as $r \to 0$ and as $r \to \infty$. The solutions of which give the behavior of $R(r)$ in these limits, $R_{r \to 0}(r)$ and $R_{r \to \infty}(r)$. The solution for all values of $r$ is found by letting $R(r)$ equal the multiplication of these solutions, times another function $F(r)$ that dictates how the solution behaves between these limits.

$$R(r) = R_{r \to 0}(r) F(r) R_{r \to \infty}(r) \qquad (i)$$

Plugging this into the original differential equation gives a new differential equation for $F(r)$. This equation is then solved, with the constraint that $R(r)$ must be normalizable, and the radial part of the wave function is attained. It is important to note that both the discreteness of energy levels and the constraint on angular momentum comes from the derivation of $F(r)$.

## Particles Orbiting A Black Hole

For the quantum mechanical analysis of gravitational collapse one is lead to the relativistic wave equations for two reasons. Primarily out of the need to express the equations in a covariant form so the equivalence principle can be utilized to show the effects of the gravitational field [1]. Secondly, to allow for the particle's motion to be relativistic. For bound state solutions in relativistic quantum mechanics the non-relativistic constraint that the particle's energy is negative (E<0) is replaced by the constraint that it is less then it's mass (E<m). To exist a wave function must be continuous and normalizable, which translates to the need for it to be finite at the $r = 0$ boundary. At the $r = r_s$ boundary the wave function must go to zero so that it can vanish for $r > r_s$ and be continuous across this boundary. This prevents the particle from having a probability of escaping the black hole to $r > r_s$. Which would violate the equivalence principle [3] as this is prohibited classically.

## Spin 0 Particles

This calculation will be done in natural units [1,5], with $G = \hbar = c = 1$. In these units the equation of motion for a spin 0 particle is given by the Klein-Gordon equation [6].

$$(\Box + m^2)\psi = 0 \tag{1}$$

In the Schwarzschild metric [7] this becomes

$$\frac{1}{1-\frac{2M}{r}} \frac{\partial^2 \psi}{\partial t^2} - \frac{1}{r^2} \frac{\partial}{\partial r}\left[r^2\left(1-\frac{2M}{r}\right)\frac{\partial \psi}{\partial r}\right] - \frac{1}{r^2 \sin\theta} \frac{\partial}{\partial \theta}\left(\sin\theta \frac{\partial \psi}{\partial \theta}\right) - \frac{1}{r^2 \sin^2\theta} \frac{\partial^2 \psi}{\partial \phi^2} + m^2 \psi = 0$$

$$\tag{2}$$

Letting,

$$\psi = R(r) Y_l^m(\theta, \phi) e^{-iEt}$$
$$j = 1 - \frac{2M}{r} \tag{3}$$

Where $Y_l^m$ are spherical harmonics, the equation becomes

$$\frac{-1}{r^2} \frac{\partial}{\partial r}\left(r^2 j \frac{\partial R}{\partial r}\right) + \left[\frac{l(l+1)}{r^2} + m^2 - \frac{E^2}{j}\right] R = 0 \tag{4}$$

After doing the following changes

$$u = rR \qquad \tilde{m} = 2Mm$$
$$x = \frac{r}{2M} \qquad \tilde{E} = 2ME \tag{5}$$

The equation transforms to the following

$$-j\frac{\partial^2 u}{\partial x^2} - \frac{1}{x^2}\frac{\partial u}{\partial x} + \left[\frac{1}{x^3} + \frac{l(l+1)}{x^2} + \tilde{m}^2 - \frac{\tilde{E}^2}{j}\right] u = 0 \tag{6}$$

This equation has two regular singular points, one at $r = 0$ ($x = 0$) and the other at $r = 2M$ ($x = 1$). As $r \to 0$ the leading terms of $\frac{\partial^2 u}{\partial x^2}$, $\frac{\partial u}{\partial x}$, and $u$ are

$$\frac{1}{x}\frac{\partial^2 u}{\partial x^2} - \frac{1}{x^2}\frac{\partial u}{\partial x} + \frac{1}{x^3} u = 0 \tag{7}$$

Multiplying the equation by $x^3$ turns it into Euler's Equation [8], with solutions

$$u = c_1 x + c_2 x \ln x \tag{8}$$

The first solution is finite as $r \to 0$, and although the second solution (for $R(r)$) diverges it is normalizable. So the wave function itself is finite as $r \to 0$.

As $r \to 2M$ ($x \to 1$, $j \to 0$) the equation behaves as

$$-j\frac{\partial^2 u}{\partial x^2} - \frac{1}{x^2}\frac{\partial u}{\partial x} - \frac{\tilde{E}^2}{j} u = 0 \tag{9}$$

Multiplying this equation by $-j$ and doing the changes of variables $b = \frac{1}{x}$, $j = 1 - b$, $y = \frac{1}{j}$ changes the equation to

$$y^3 \frac{\partial^2 u}{\partial y^2} + 2(y^2 + 1)\frac{\partial u}{\partial y} + \tilde{E}^2 u = 0 \tag{10}$$

As $r \to 2M$, $x \to 1$, $b \to 1$, $j \to 0$, $y \to \infty$, such that the term $y^2 + 1 \approx y^2$ and the equation behaves as

$$y^2 \frac{\partial^2 u}{\partial y^2} + 2y \frac{\partial u}{\partial y} + \frac{\tilde{E}^2}{y} u = 0 \qquad (11)$$

Which is the Bessel Equation (transformed) [8] with solutions

$$u = y^{-\frac{1}{2}} \left[ c_1 J_{-1}\left(-2\tilde{E} y^{-\frac{1}{2}}\right) + c_2 Y_{-1}\left(-2\tilde{E} y^{-\frac{1}{2}}\right) \right] \qquad (12)$$

Where $J$ and $Y$ are the ordinary Bessel functions of the first and second kind. The argument inside the Bessel functions goes to zero as $y \to \infty$, forcing us to drop the $c_2$ term because it diverges. Therefore in the limit as $r \to 2M$ ($y \to \infty$) the solution goes as

$$\lim_{y \to \infty} u = \lim_{y \to \infty} c_1 y^{-\frac{1}{2}} J_{-1}\left(-2\tilde{E} y^{-\frac{1}{2}}\right) \approx \lim_{y \to \infty} c_1 \frac{2\tilde{E}}{y} = 0 \qquad (13)$$

Showing that the wave function goes to zero at $r = r_s$ and the boundary condition is satisfied.

## Spin ½ Particles

The wave equation for a spin ½ particle is given by the Dirac equation.

$$(i\gamma^\mu \partial_\mu - m)\psi = 0 \qquad (14)$$

The gamma matrices satisfy the equation [5]

$$\{\gamma^\mu, \gamma^\nu\} = 2Ig^{\mu\nu} \qquad (15)$$

Where $I$ is the identity matrix. For the Schwarzschild metric the gamma matrices are given by

$$\gamma^0 = \frac{1}{\sqrt{1 - \frac{2M}{r}}} \begin{pmatrix} 1 & 0 \\ 0 & -1 \end{pmatrix} \qquad \gamma^\theta = \frac{1}{r} \begin{pmatrix} 0 & \sigma_\theta \\ -\sigma_\theta & 0 \end{pmatrix}$$

$$\gamma^r = \sqrt{1 - \frac{2M}{r}} \begin{pmatrix} 0 & \sigma_r \\ -\sigma_r & 0 \end{pmatrix} \qquad \gamma^\phi = \frac{1}{r \sin\theta} \begin{pmatrix} 0 & \sigma_\phi \\ -\sigma_\phi & 0 \end{pmatrix} \qquad (16)$$

where

$$\sigma_r = \sin\theta\cos\phi\sigma_1 + \sin\theta\sin\phi\sigma_2 + \cos\theta\sigma_3$$
$$\sigma_\theta = \cos\theta\cos\phi\sigma_1 + \cos\theta\sin\phi\sigma_2 - \sin\theta\sigma_3 \quad (17)$$
$$\sigma_\phi = -\sin\phi\sigma_1 + \cos\phi\sigma_2$$

and $\sigma_1, \sigma_2, \sigma_3$ are the Pauli Spin Matrices. Defining

$$w = \sqrt{1 - \frac{2M}{r}}$$
$$\psi = \begin{pmatrix} F(\vec{r})e^{-iEt} \\ G(\vec{r})e^{-iEt} \end{pmatrix} \quad (18)$$

The equation becomes

$$\frac{E}{w}\begin{pmatrix} 1 & 0 \\ 0 & -1 \end{pmatrix}\begin{pmatrix} F \\ G \end{pmatrix} + iw\begin{pmatrix} 0 & \sigma_r \\ -\sigma_r & 0 \end{pmatrix}\begin{pmatrix} F_{,r} \\ G_{,r} \end{pmatrix} + \frac{i}{r}\begin{pmatrix} 0 & \sigma_\theta \\ -\sigma_\theta & 0 \end{pmatrix}\begin{pmatrix} F_{,\theta} \\ G_{,\theta} \end{pmatrix}$$
$$+ \frac{i}{r\sin\theta}\begin{pmatrix} 0 & \sigma_\phi \\ -\sigma_\phi & 0 \end{pmatrix}\begin{pmatrix} F_{,\phi} \\ G_{,\phi} \end{pmatrix} - m\begin{pmatrix} F \\ G \end{pmatrix} = 0 \quad (19)$$

Which can be rewritten as

$$\frac{EF}{w} + i\tilde{\vec{\sigma}}\cdot\vec{\nabla}G - mF = 0$$
$$\frac{-EG}{w} - i\tilde{\vec{\sigma}}\cdot\vec{\nabla}F - mG = 0 \quad (20)$$

Where $\tilde{\vec{\sigma}} = (w\sigma_r, \sigma_\theta, \sigma_\phi)$. The curvature of space-time only affects the radial term in the dot product, hinting that the $\tilde{\vec{\sigma}}\cdot\vec{\nabla}$ should be broken up into its radial and orbital terms

$$i\tilde{\vec{\sigma}}\cdot\vec{\nabla} = iw\vec{\sigma}\cdot\hat{r}\frac{\partial}{\partial r} + i\left(\vec{\sigma}\cdot\vec{\nabla} - \vec{\sigma}\cdot\hat{r}\frac{\partial}{\partial r}\right) \quad (21)$$

It can be shown that the orbital term can be rewritten as [6]

$$i\left(\vec{\sigma}\cdot\vec{\nabla} - \vec{\sigma}\cdot\hat{r}\frac{\partial}{\partial r}\right) = i\vec{\sigma}\cdot\hat{r}\frac{\vec{\sigma}\cdot\vec{L}}{r} \quad (22)$$

Breaking $F$ and $G$ into their radial and orbital terms

$$F = f(r)\mathbf{Y}_{jm}^{k}(\theta,\phi)$$
$$G = g(r)\mathbf{Y}_{jm}^{-k}(\theta,\phi) \tag{23}$$

Where $\mathbf{Y}_{jm}^{k}$ are the generalized spherical harmonics [6] defined as

$$\mathbf{Y}_{jm}^{k} = -\operatorname{sgn} k \sqrt{\frac{k+\frac{1}{2}-m}{2k+1}}\alpha Y_{l,m-\frac{1}{2}} + \sqrt{\frac{k+\frac{1}{2}+m}{2k+1}}\beta Y_{l,m+\frac{1}{2}} \tag{24a}$$

$$\alpha = \begin{pmatrix} 1 \\ 0 \end{pmatrix} \qquad k = \pm\left(j+\frac{1}{2}\right) \tag{24b}$$

$$\beta = \begin{pmatrix} 0 \\ 1 \end{pmatrix}$$

Where $Y_{l,m}$ are spherical harmonics. $k$ is positive for $l = j+\frac{1}{2}$ (so $k = l$) and negative for $l = j-\frac{1}{2}$ (so $k = -(l+1)$). And $\operatorname{sgn} k$ is $+1$ if $k > 0$ and $-1$ for $k < 0$.
The generalized spherical harmonics have the properties

$$\vec{\sigma}\cdot\hat{r}\mathbf{Y}_{jm}^{k}(\hat{r}) = -\mathbf{Y}_{jm}^{-k}(\hat{r})$$
$$\vec{\sigma}\cdot\vec{L}\mathbf{Y}_{jm}^{k}(\hat{r}) = -(k+1)\mathbf{Y}_{jm}^{k}(\hat{r}) \tag{25}$$

With this Equation 20 can be rewritten as

$$\left(\frac{E}{w}-m\right)f + w\frac{\partial g}{\partial r} + \frac{1-k}{r}g = 0 \tag{26}$$

$$\left(\frac{E}{w}+m\right)g - w\frac{\partial f}{\partial r} - \frac{1+k}{r}f \tag{27}$$

Changing variables to $x = \dfrac{r}{2M}, b = \dfrac{1}{x}, w = \sqrt{1-b}$ and defining the constants $\tilde{E} = 2ME$ and $\tilde{m} = 2Mm$ gives the equations the form

$$\left(\frac{\tilde{E}}{w}-\tilde{m}\right)f + \frac{1}{2}(1-w^2)^2\frac{\partial g}{\partial w} + (1-k)(1-w^2)g = 0 \tag{26'}$$

$$\left(\frac{\tilde{E}}{w}+\tilde{m}\right)g - \frac{1}{2}(1-w^2)^2 \frac{\partial f}{\partial w} - (1+k)(1-w^2)f = 0 \tag{27'}$$

Solving for $f$ in Equation 26', differentiating, and plugging the expressions for $f$ and $\frac{\partial f}{\partial w}$ in terms of $g, \frac{\partial g}{\partial w}, \frac{\partial^2 g}{\partial w^2}$ into Equation 27' (and vice versa for $g$ and Equation 26') the differential equations for $f$ and $g$ become (letting $B_\pm = \frac{\tilde{E}}{w} \pm \tilde{m}$)

$$B_+ B_- f + \frac{1}{4}(1-w^2)^2 \left[ -4w(1-w^2)\frac{\partial f}{\partial w} + \frac{\tilde{E}(1-w^2)^2}{w^2 B_+}\frac{\partial f}{\partial w} + (1-w^2)^2 \frac{\partial^2 f}{\partial w^2} \right]$$

$$+ \frac{1}{2}(1+k)(1-w^2)^2 \left[ -2wf + \frac{\tilde{E}(1-w^2)}{w^2 B_+}f + (1-w^2)\frac{\partial f}{\partial w} \right] \tag{28}$$

$$+ (1-k)(1-w^2)\left[ \frac{1}{2}(1-w^2)^2 \frac{\partial f}{\partial w} + (1+k)(1-w^2)f \right] = 0$$

$$B_+ B_- g + \frac{1}{4}(1-w^2)^2 \left[ -4w(1-w^2)\frac{\partial g}{\partial w} + \frac{\tilde{E}(1-w^2)^2}{w^2 B_-}\frac{\partial g}{\partial w} + (1-w^2)^2 \frac{\partial^2 g}{\partial w^2} \right]$$

$$+ \frac{1}{2}(1-k)(1-w^2)^2 \left[ -2wg + \frac{\tilde{E}(1-w^2)}{w^2 B_-}g + (1-w^2)\frac{\partial g}{\partial w} \right] \tag{29}$$

$$+ (1+k)(1-w^2)\left[ \frac{1}{2}(1-w^2)^2 \frac{\partial g}{\partial w} + (1-k)(1-w^2)g \right] = 0$$

As with the spin 0 case these equations have two regular singular points, one at $r=0$ ($w=i\infty$) and the other at $r=2M$ ($w=0$). As $r \to 2M$ ($w \to 0$) Equation 28 takes the form

$$\frac{\tilde{E}^2}{w^2}f + \frac{1}{4w}\frac{\partial f}{\partial w} + \frac{1}{4}\frac{\partial^2 f}{\partial w^2} + \frac{(1+k)}{2w}f = 0 \tag{30}$$

Equation 29 taking the same form with the replacements $f \to g$ and $k \to -k$. Multiplying by $4w^2$ changes the equation to

$$w^2 \frac{\partial^2 f}{\partial w^2} + w \frac{\partial f}{\partial w} + \left[ 2(1+k)w + \tilde{E}^2 \right] f = 0 \qquad (31)$$

Which is again the Bessel Equation (transformed), with solutions

$$f = c_1 J_{2\tilde{E}}\left(2\sqrt{2(1+k)w}\right) + c_2 Y_{2\tilde{E}}\left(2\sqrt{2(1+k)w}\right) \qquad (32)$$

Where $c_2$ must be set to zero because this term diverges as $r \to 2M$ ($w \to 0$). In this limit $f$ (and $g$) goes to

$$\lim_{w \to 0} f = \lim_{w \to 0} c_1 J_{2\tilde{E}}\left(2\sqrt{2(1+k)w}\right) \approx \lim_{w \to 0} c_1 \frac{[2(1+k)w]^{\tilde{E}}}{\Gamma\left(2\tilde{E}+1\right)} = 0 \qquad (33)$$

Showing that the wave function goes to zero at $r = r_s$ and the boundary condition is satisfied. As $r \to 0$ ($w \to i\infty$) Equation 28 takes the form

$$\frac{1}{4} w^8 \frac{\partial^2 f}{\partial w^2} + w^7 \frac{\partial f}{\partial w} - (1+k)w^5 f = 0 \qquad (34)$$

With Equation 29 taking the same form with the replacements $f \to g$ and $k \to -k$. Multiplying by $\frac{4}{w^6}$ gives the equation the form

$$w^2 \frac{\partial^2 f}{\partial w^2} + 4w \frac{\partial f}{\partial w} - 4(1+k)b^{-1} f = 0 \qquad (35)$$

Which is again the Bessel Equation (transformed) with solutions

$$f = w^{-\frac{3}{2}}\left[ c_1 J_{-3}\left(-4i\sqrt{\frac{1+k}{w}}\right) + c_2 Y_{-3}\left(-4i\sqrt{\frac{1+k}{w}}\right) \right] \qquad (36)$$

Once again the $c_2$ term diverges as $r \to 0$ ($w \to i\infty$) and must be set to zero. In this limit $f$ (and $g$) behaves as

$$\lim_{w \to i\infty} f = \lim_{w \to i\infty} c_1 w^{-\frac{3}{2}} J_{-3}\left(-4i\sqrt{\frac{1+k}{w}}\right) \approx \lim_{w \to i\infty} c_1 \frac{\left[2i(1+k)^{\frac{1}{2}}\right]^3}{\Gamma(4)w^3} = 0 \qquad (37)$$

Showing that the wave function is finite as $r \to 0$.

## Conclusion

With this, it can be concluded that bound state wave functions exist for spin 0 and spin ½ particles inside the schwarzschild radius of a black hole.  Showing that the particles are not confined to an infinitesimal point.  As with the case of electric collapse, if an electron cannot spiral into a nucleus, then neither does this happen in plasma consisting of many electrons and nuclei.  Showing that after undergoing gravitational collapse, like plasma, matter will not contract to an infinitesimal point.  That quantum effects prevents singularity formation.  To finish the calculation for the bound state wave functions, by finding the function that dictates how the wave functions vary between the limits as $r \to 0$ and $r \to 2M$, is unwarranted since the wave equation itself (other then the knowledge that singularities do not form) is not applicable to nature.  Just as the hydrogen atom wave equation is not applicable in any way to plasma physics.

The author would like to thank Douglas MacLaughlin and Gerard LaVarnway for discussions on the properties of analytic functions, and Leonid Pryadko for discussions on the limits of functions to complex powers.